\documentclass[aps,prd,twocolumn,amsmath,showpacs,superscriptaddress,
nofootinbib,
nopreprintnumbers]{revtex4-1}

\usepackage{verbatim}
\usepackage[T1]{fontenc}
\usepackage[utf8]{inputenc}
\usepackage[american]{babel}
\usepackage{epsfig}
\usepackage{graphicx}
\usepackage{booktabs}
\usepackage{multirow}
\usepackage{dcolumn}
\usepackage{amsmath}
\usepackage{mathtools}
\usepackage{amsfonts}
\usepackage{amssymb}
\usepackage{epstopdf}
\usepackage{bm}
\usepackage{siunitx}
\usepackage{braket}
\usepackage{enumitem}
\usepackage{soul}
\usepackage{booktabs}
\usepackage[table]{xcolor}
\usepackage{color}
\usepackage{transparent}
\usepackage{pifont}
\usepackage[normalem]{ulem}
\usepackage{pbox}

\definecolor{navyblue}{rgb}{0.0, 0.0, 0.5}
\definecolor{royalblue}{rgb}{0.25, 0.41, 0.88}
\definecolor{cadmiumgreen}{rgb}{0.0, 0.42, 0.24}
\definecolor{blue-violet}{rgb}{0.54, 0.17, 0.89}
\definecolor{darkviolet}{rgb}{0.58, 0.0, 0.83}
\definecolor{orange(colorwheel)}{rgb}{1.0, 0.5, 0.0}

\usepackage{hyperref}
\hypersetup{
    colorlinks=true, % false: boxed links; true: colored links
    linkcolor=royalblue, % color of internal links (change box color with linkbordercolor)
    citecolor=magenta}
\usepackage[capitalize]{cleveref}
\usepackage{multirow}

\renewcommand\[{\left[}

\usepackage{booktabs}
\usepackage{multirow}
\usepackage{dcolumn}
\usepackage{colortbl}

\definecolor{magenta(process)}{rgb}{1.0, 0.0, 0.56}

\definecolor{darkspringgreen}{rgb}{0.09, 0.45, 0.27}

\definecolor{royalblue(web)}{rgb}{0.25, 0.41, 0.88}
\begin{document}

\title{Non-standard neutrino cosmology dilutes the lensing anomaly
} 
\author{Ivan Esteban}
\email{esteban.6@osu.edu}
\affiliation{Center for Cosmology and AstroParticle Physics (CCAPP), Ohio State University, Columbus, OH 43210}
\affiliation{Department of Physics, Ohio State University, Columbus, OH 43210}
\author{Olga Mena}
\email{omena@ific.uv.es }
\affiliation{IFIC, Universidad de Valencia-CSIC, 46071, Valencia, Spain}
\author{Jordi Salvado}
\email{jsalvado@icc.ub.edu}
\affiliation{Departament de Física Quàntica i Astrofísica and Institut de Ciencies del Cosmos, Universitat de Barcelona, Diagonal 647, E-08028 Barcelona, Spain}

\date{\today}

\preprint{}
\begin{abstract}
Despite the impressive success of the standard cosmological model, several anomalies defy its triumph. Among them is the so-called lensing anomaly: the Planck satellite observes stronger CMB gravitational lensing than expected. The role of neutrinos in this anomaly has been mostly overlooked, despite their key role in CMB lensing, because in the standard scenario they tend to increase the tension. Here, we show that this strongly depends on the assumed neutrino equation of state. We demonstrate that if neutrinos have yet undiscovered long-range interactions, the lensing pattern is significantly affected, rendering the lensing anomaly as a statistical fluctuation. Our results thus open up a window to link anomalous CMB lensing with present and future cosmological, astrophysical, and laboratory measurements of neutrino properties.

\end{abstract}
%%%%%%%%%%%%%%%%%%%%%%%%%%%%%%%%%%%%%%%%%%%%%%%%%%%%%%%%%%%%%%%%%
\maketitle
%%%%%%%%%%%%%%%%%%%%%%%%%%%%%%%%%%%%%%%%%%%%%%%%%%%%%%%%%%%%%%%%%%
\section{Introduction} \label{sec:intro} 

The standard cosmological model, remarkably simple but with profound implications, remains strikingly successful in many diverse environments~\cite{Beutler:2011hx,Ross:2014qpa,Riess:2016jrr,Alam:2016hwk,Pan-STARRS1:2017jku,DES:2018paw,Planck:2018vyg,Agathe:2019vsu,Blomqvist:2019rah,DES:2021wwk,ACT:2020gnv}. The precision is such, that self-consistency tests are feasible and accurate. These tests check not only if individual datasets are consistent with the model, but also whether the same model \emph{simultaneously} explains all observations or if, on the contrary, new ingredients are needed.

It is in these consistency tests where the standard model may meet its match.
Indeed, there exist several anomalies that can not be fully understood in the minimal cosmological constant plus cold dark matter ($\Lambda$CDM) scenario. The most statistically significant one is the discrepancy of the Hubble constant extracted from nearby universe probes and that derived from Cosmic Microwave Background (CMB) measurements~\cite{DiValentino:2021izs,Schoneberg:2021qvd,DiValentino:2020zio}. Another one is related to the clustering parameter $\sigma_8$, whose values differ for CMB and weak lensing estimates~\cite{DiValentino:2020vvd}. The third one, the so-called lensing anomaly~\cite{Planck:2018vyg,Motloch:2018pjy}, is the focus of this work.

CMB anisotropies get blurred due to gravitational lensing by the large scale structure of the Universe: photons from different directions are mixed and the peaks at large multipoles are smoothed. The amount of lensing is a precise prediction of the $\Lambda$CDM model: the consistency of the model can be checked by artificially increasing lensing by a factor $A_{\rm{lens}}$~\cite{Calabrese:2008rt} (\emph{a priori} an unphysical parameter). If $\Lambda$CDM consistently describes all CMB data, observations should prefer $A_{\rm{lens}}=1$.

Intriguingly, Planck CMB data shows a \emph{preference} for additional lensing. Indeed, the reference analysis of temperature and polarization anisotropies suggest $A_\mathrm{lens} > 1$ at 3$\sigma$. CMB lensing also introduces a non-trivial four-point correlation function, and therefore it can be independently measured. Adding this information somewhat diminishes the tension, albeit the value of the lensing amplitude is still above the canonical one by about $2\sigma$.
The lensing anomaly is robust against changes in the foreground modeling in the baseline likelihood, and was already discussed in previous data releases, although it is currently more significant due to the lower reionization optical depth preferred by the Planck 2018 data release. A recent result from the Atacama Cosmology Telescope is compatible with $A_\mathrm{lens}=1$~\cite{ACT:2020gnv}, but the results are consistent with Planck within uncertainties. 

Barring systematic errors or a rare statistical fluctuation,  the lensing anomaly could have its origin in new physics scenarios. Attempts in the literature have changed either the geometrical or the gravitational sectors of the theory. A closed universe~\cite{DiValentino:2019qzk} has been shown to solve the internal tensions in Planck concerning the cosmological parameter values at different angular scales, alleviating the $A_{\rm{lens}}$ anomaly. However, the positive curvature scenario is strongly rejected by other cosmological observations such as  Baryon Acoustic Oscillations (BAO)~\cite{Efstathiou:2020wem}. Modified gravity theories or a modified primordial perturbation spectrum have also been considered as possible scenarios where to solve the $A_{\rm{lens}}$ anomaly~\cite{DiValentino:2015bja,Moshafi:2020rkq,Domenech:2019cyh,Domenech:2020qay,Hazra:2022rdl}, but it is also unclear if BAO measurements are compatible with these solutions~\cite{Domenech:2020qay}.  
 
Here, we show that non-standard neutrino properties can lead to unexpected lensing and dilute the preference for $A_\mathrm{lens} \neq 1$. Neutrinos are key to structure formation in our Universe, but due to a competition of different effects, in the standard paradigm they \emph{reduce} CMB lensing~\cite{Kaplinghat:2003bh,Lesgourgues:2005yv}. However, we demonstrate that this dramatically depends on the assumed equation of state for the cosmological neutrino fluid. Under the presence of long-range neutrino interactions~\cite{Esteban:2021ozz}, largely allowed by other constraints~\cite{Lessa:2007up,Blinov:2019gcj,Pasquini:2015fjv,Agostini:2015nwa,Blum:2018ljv,Brune:2018sab,Escudero:2019gfk,Brdar:2020nbj}, the generally assumed ideal gas equation of state no longer holds. This pivotal difference makes neutrinos \emph{increase} CMB lensing, alleviating the $A_{\rm{lens}}$ anomaly. Crucially, other cosmological datasets such as BAO do not compromise this effect. 

 The structure of the manuscript is as follows. We start in Sec.~\ref{sec:nu} by describing the neutrino interaction model and its impact on the equation of state. In \cref{sec:observables}, we show how different cosmological observables are affected, describing in detail the effect on CMB lensing. Section \ref{sec:method} presents our quantitative results on the anomaly. We draw our conclusions and highlight future directions in Sec.~\ref{sec:concl}.

%%%%%%%%%%%%%%%%%%%%%%%%%%%% 
\section{Long-range neutrino interactions }
%%%%%%%%%%%%%%%%%%%%%%%%%%%% 
\label{sec:nu}

We consider neutrino self-interactions mediated by a very light scalar field $\phi$ with mass $M_\phi$, so that neutrinos source a classical field making long-range interaction effects relevant~\cite{Fardon:2003eh,Kaplan:2004dq,Gu:2003er,Peccei:2004sz,Bean:2000zm,Bean:2001ys,Esteban:2021ozz}. The corresponding interaction Lagrangian is
\begin{equation}
\mathcal{L}_\mathrm{int} = - g \phi \bar{\nu} \nu \, ,
\label{eq:Lagrangian}
\end{equation}
the standard scenario corresponding to $g=0$. For simplicity, we consider one scalar field universally coupled to all three neutrino mass eigenstates, that we assume to have a common mass $m_\nu$ (cosmological effects of neutrino mass splittings are very challenging to detect~\cite{Lesgourgues:2006nd,Pritchard:2008wy,Jimenez:2010ev}).

The phenomenology depends on the parameter values (more details are given in Ref.~\cite{Esteban:2021ozz}). For $M_\phi \ll m_\nu$, the interaction range is larger than interneutrino distances, and long-range effects are relevant as long as the scalar field is dynamical. This requires $M_\phi \gg H$ (with $H$ the Hubble parameter) at all relevant times, which for CMB observables corresponds to ${M_\phi \gtrsim 10^{-25} \, \mathrm{eV}}$.  Long-range effects are governed by the ratio $g m_\nu/M_\phi$. The main constraint, from the observation of neutrino oscillations, is $g m_\nu/M_\phi \lesssim 10^5$ (although this can be evaded if the interaction has flavor structure). Scattering effects depend on $g$ for light mediators, so they can be small even if long-range effects are large (and viceversa). An example of models with extra scattering are those invoked to alleviate the Hubble tension~\cite{Hannestad:2005ex,Archidiacono:2014nda,Cyr-Racine:2013jua,Lancaster:2017ksf,Blinov:2019gcj,Park:2019ibn,Kreisch:2019yzn,Escudero:2019gvw,RoyChoudhury:2020dmd,Brinckmann:2020bcn,Ghosh:2021axu}. SN 1987A and other laboratory constraints~\cite{Lessa:2007up,Blinov:2019gcj,Pasquini:2015fjv,Agostini:2015nwa,Blum:2018ljv,Brune:2018sab,Escudero:2019gfk,Brdar:2020nbj} require $g \lesssim 10^{-7}$. Since we are interested in long-range effects, we focus on the regime where the coupling is tiny, $g \ll 10^{-7}$. This guarantees that scatterings can be neglected and that the scalar field is never produced on-shell (so $N_\mathrm{eff}$ does not change from its standard value~\cite{Huang:2017egl}).

The impact of long-range interactions on cosmic neutrinos is simple (see Ref.~\cite{Esteban:2021ozz} for more details): when the density is large and the typical energy is of the order of the neutrino mass, neutrinos source a classical scalar field. This, in turn, induces an effective neutrino mass $\tilde{m} = m_\nu + g \phi \neq m_\nu$ via the Lagrangian, see Eq.~(\ref{eq:Lagrangian}). The energy and pressure of the scalar field are also relevant, and they modify the equation of state. Finally, when neutrinos become non-relativistic, the scalar field induces an attractive force among them. This abruptly condenses all cosmic neutrinos into lumps of size $\lesssim M_\phi^{-1}$~\cite{Afshordi:2005ym, Beca:2005gc,Kaplinghat:2006jk,Bjaelde:2007ki,Bean:2007nx,Gao:2021fyk,Smirnov:2022sfo}. 

\begin{figure}[t]
\begin{center}
\includegraphics[width=\columnwidth]{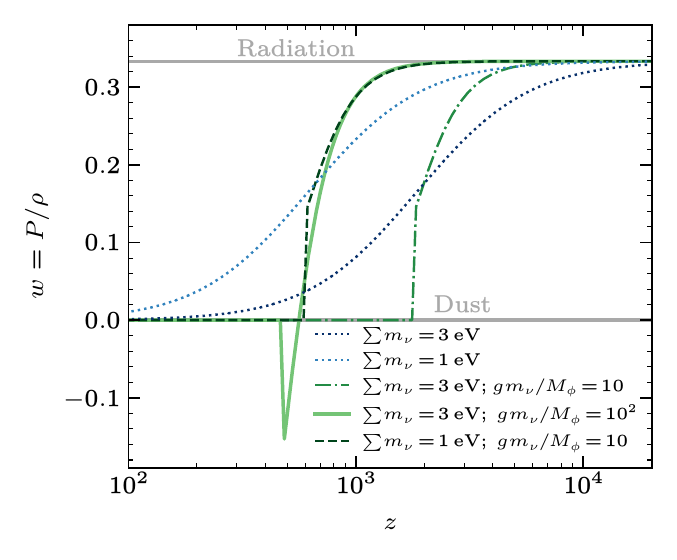} 
\caption{Modified neutrino equation of state $w$ due to long-range interactions (see Ref.~\cite{Esteban:2021ozz} for the numerical computation). As $w$ controls the rate of neutrino energy density dilution, $\dot{\rho} / \rho = -3 H(1+w)$, \emph{the neutrino energy at different redshifts is dramatically affected, which directly impacts many cosmological observables}.}
\label{fig:w}
\end{center}
\end{figure}

\Cref{fig:w} summarizes how long-range interactions affect the evolution of cosmic neutrinos by changing their equation of state (defined as the ratio between pressure and energy density) as a function of redshift $z$. Under the presence of a long-range interaction, neutrinos behave as radiation for longer times (because $\tilde{m} < m_\nu$ due to scalar interactions being attractive). Later, the negative pressure of the scalar field promptly changes the equation of state to that of dust. This effect is further enhanced by the abrupt formation of lumps. 

\section{Impact on CMB observables}
\label{sec:observables}
The modified equation of state directly impacts the precisely measured CMB anisotropies. As we shall show below, it also increases the observed lensing.

\begin{figure}[t]
\begin{center}
\includegraphics[width=\columnwidth]{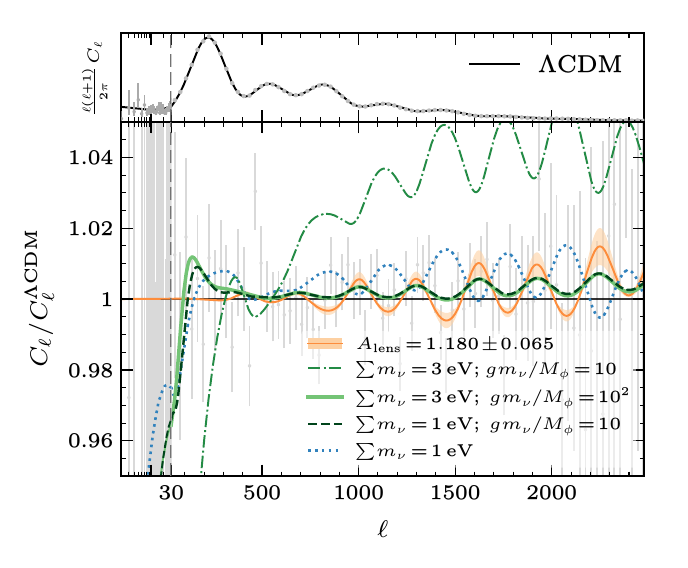} 
\caption{
Modified CMB temperature anisotropies by massive and interacting neutrinos, together with the prediction for $A_{\rm{lens}} \neq 1$. 
\emph{Neutrino long-range interactions mimic an enhanced lensing contribution, as preferred by the data}.
}
\label{fig:cmb}
\end{center}
\end{figure}

\Cref{fig:cmb} shows how neutrino masses and interactions impact CMB temperature anisotropies. The bottom panel shows the anisotropy ratios to the canonical $\Lambda$CDM scenario with massless neutrinos, together with Planck data. In this figure, we have fixed the cosmological parameters $\{\theta_s, \omega_b, \omega_\mathrm{cdm}, A_s, n_s, \tau_\mathrm{reio}\}$.

The main effects of long-range interactions arise due to the Integrated Sachs Wolfe effect (for ${\ell \lesssim 500}$), directly sensitive to the neutrino equation of state; and a reduced Silk damping scale when the neutrino energy density redshifts as dust (visible as an enhancement for large $\ell$). These are also the standard effects of neutrino masses on CMB temperature anisotropies~\cite{Lesgourgues:2018ncw,Esteban:2021ozz}, but since they can be traced back to how the neutrino equation of state gets modified when they become non-relativistic, they are directly impacted by long-range interactions. As a consequence, strong long-range interactions completely remove the CMB bound on neutrino masses~\cite{Esteban:2021ozz}.

Interestingly, we also observe additional wiggles at large $\ell$, that mimic the scenario with artificially modified lensing $A_\mathrm{lens} \neq 1$. The effect of non-interacting massive neutrinos is out of phase with $A_\mathrm{lens} > 1$, indicating that neutrino masses typically \emph{reduce} CMB lensing as we describe below. However, as we show below, introducing a long-range interaction \emph{increases} the lensing. This effect is key to our results.

Indeed, the orange shaded region in \cref{fig:cmb} corresponds to the massless neutrino case with the value of $A_\mathrm{lens}$ inferred from a fit to Planck temperature and polarization measurements~\cite{Planck:2018vyg}. As we see, it lies very close to the interacting neutrino scenario. However, it is not fully degenerate as long-range interactions introduce additional effects (particularly a reduced Silk damping). This allows future observations to disentangle long-range interactions from other solutions to the lensing anomaly.

\begin{figure}[t]
\begin{center}
\includegraphics[width=\columnwidth]{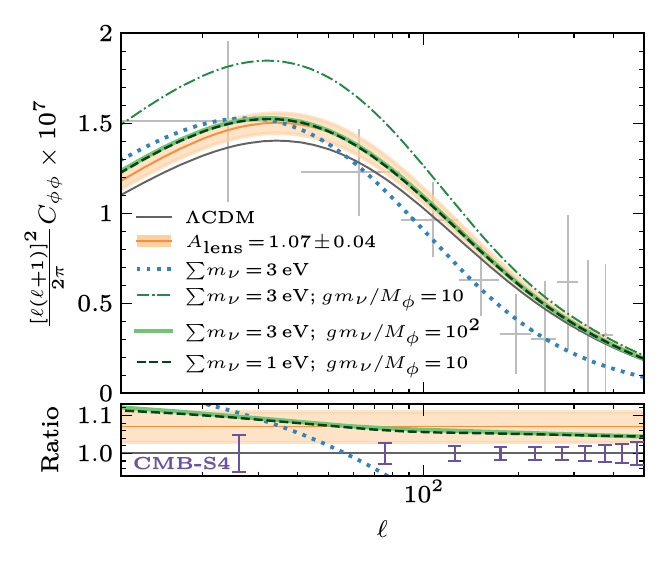} 
\caption{Modified CMB lensing power spectrum by massive and interacting neutrinos, and the prediction for $A_\mathrm{lens} \neq 1$. 
%The bottom panel shows the ratios to the canonical $\Lambda$CDM scenario, together with the expected error bars for the next-generation ground-based CMB experiment, CMB-S4. 
Unlike neutrino masses, \emph{neutrino long-range interactions enhance lensing, closely mimicking $A_\mathrm{lens} > 1$}. Future observations could discriminate among these two effects (see the main text for more details).}
\label{fig:lensing}
\end{center}
\end{figure}

\Cref{fig:lensing} shows how neutrino masses and interactions affect the CMB lensing pattern. The bottom panel shows the ratios to the canonical $\Lambda$CDM scenario with massless neutrinos. The data points in the upper plot correspond to Planck~\cite{Planck:2018lbu}, while in the bottom panel they refer to the errors expected by the next-generation ground-based CMB experiment CMB-S4~\cite{CMB-S4:2016ple,Brinckmann:2018owf}. The orange band shows the enhanced lensing preferred by a fit to Planck CMB temperature, polarization and lensing~\cite{Planck:2018vyg}.

In the standard scenario, neutrinos reduce CMB lensing because they are hot thermal relics with a very large velocity dispersion, and therefore they suppress clustering at small scales. Furthermore, when they become non-relativistic their energy density dilutes slower, enhancing the expansion of the Universe and further suppressing structure formation. However, they also have lensing-enhancing effects. On the one hand, they cluster as matter at large scales, enhancing density perturbations and thus the lensing. %N.B.: if you look at any plot of the matter power spectrum, you'll see that neutrino masses have no effect at small k. But this is plotting the power spectrum of the *relative* density perturbations (delta rho/rho). Lensing is sensitive to the gravitational potential, i.e., directly to delta rho. And if rho is larger delta rho is also larger
On the other hand, their modified equation of state before recombination would affect the well-measured CMB sound horizon, which must be compensated by slightly decreasing $H_0$, and a slower expansion rate of the Universe also enhances structure formation. Without long-range interactions (dotted line in \cref{fig:lensing}), the last two effects are subdominant, particularly at large $\ell$. Overall, neutrino masses tend to \emph{reduce} CMB lensing.

However, long-range interactions flip this effect. The relativistic to non-relativistic transition is more abrupt: neutrinos move close to the speed of light for a shorter time (c.f.~\cref{fig:w}). The energy density in neutrinos is also smaller as scalar interactions are attractive~\cite{Esteban:2021ozz}. Therefore, the physical effects that suppress CMB lensing are diminished; and those enhancing it are increased. The overall effect is now the opposite: neutrino masses and interactions \emph{enhance} the lensing power spectrum.

Intriguingly, we note from \cref{fig:lensing} that the effect of massive neutrinos with $\sum m_\nu=3$~eV and $g m_\nu/M_\phi=10^2$ is almost degenerate with the massless neutrino case with $A_{\rm{lens}}=1.07 \pm 0.04$. Given current uncertainties, it is always possible to tune the neutrino mass and the interaction strength to produce a CMB lensing pattern very similar to that obtained with
$A_\mathrm{lens}\neq 1$. Since the degeneracy is not perfect,  next-generation CMB observations, as those from CMB-S4, might disentangle between these two effects, as  the error bars in the bottom panel of Fig.~\ref{fig:lensing} show~\cite{Renzi:2017cbg}.

The presence of a long-range interaction in the neutrino sector thus generates very rich physics. Its effects turn on at a specific time (when neutrino masses start to be relevant), and make neutrinos behave as radiation even if they have \emph{large} masses, at the same time enhancing CMB lensing as neutrinos increase perturbation growth in a different way than they would do in the non-interacting case. This makes long-range neutrino interactions a very attractive and predictive scheme to solve the $A_{\rm{lens}}$ anomaly.

\begin{figure}[b]
\begin{center}
\includegraphics[width=\columnwidth]{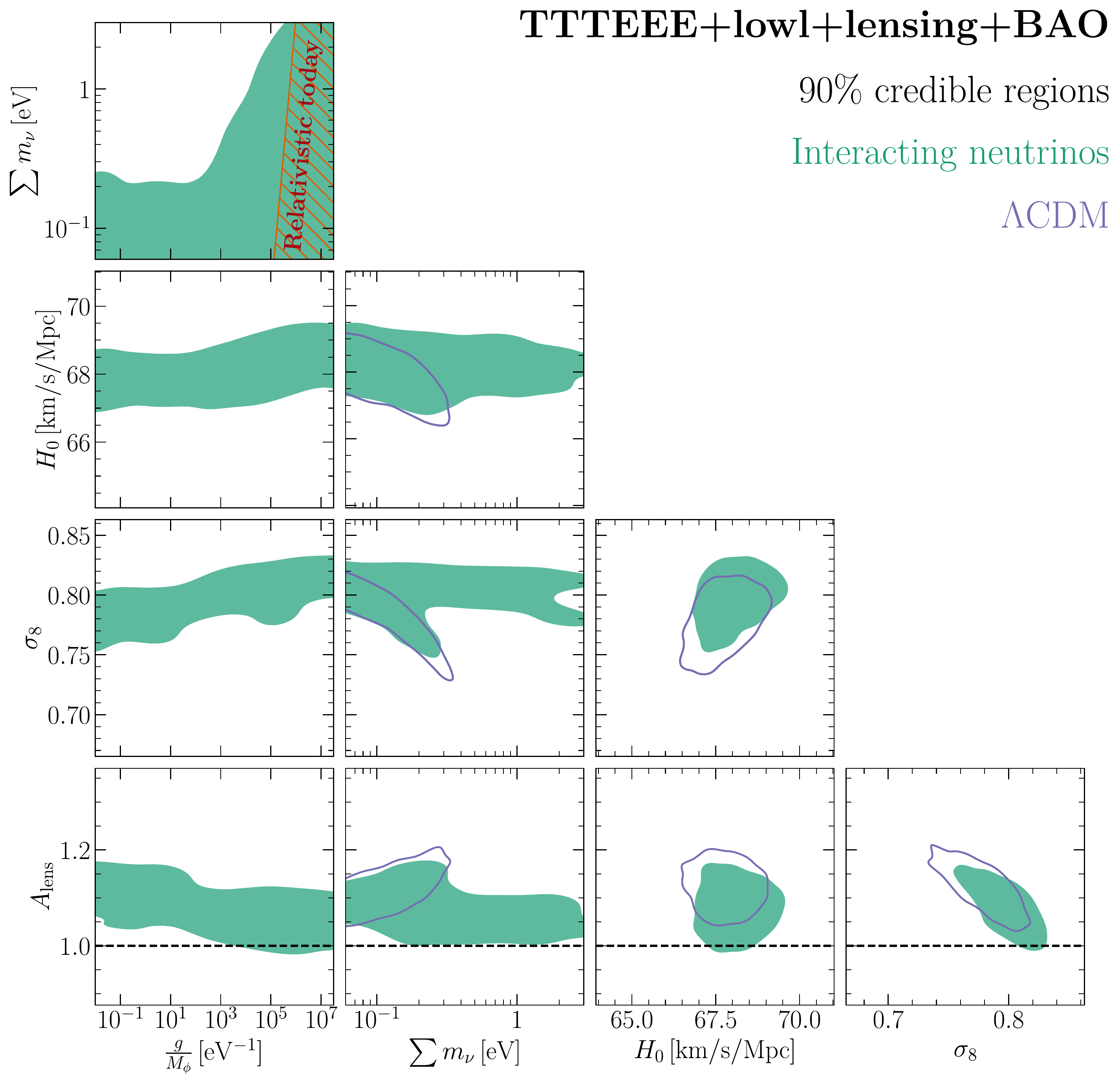} 
\caption{Allowed parameter values and correlations after including all relevant datasets. The significance of the $A_\mathrm{lens}$ anomaly as read from this figure is prior-dependent and subject to Bayesian volume effects (see text), hence we do \emph{not} draw our main quantitative conclusions from here. \emph{Allowing for neutrino long-range interactions dilutes the anomalous preference for $A_{\rm lens} > 1$. The shift in other parameters is mild}.}
\label{fig:triangle}
\end{center}
\end{figure}

\begin{table*}[t]
\begin{center}
\begin{tabular}{c@{\hskip 0.5cm}c@{\hskip 0.5cm}c@{\hskip 0.5cm}c}
\multicolumn{4}{c}{$ \boldsymbol{\Delta \chi^2_\mathrm{eff} = \chi^2_\mathrm{eff}(A_\mathrm{lens}=1) - \chi^2_\mathrm{eff}(A_\mathrm{lens}\neq 1)}$}\\[0.2cm]
\toprule
& TTTEEE+lowl & TTTEEE+lowl+lensing & TTTEEE+lowl+lensing+BAO\\
\midrule
\pbox{20cm}{\texttt{Plik} (reference):  $\Lambda$CDM} & \pbox{20cm}{9.66 \\ \small{$p=0.2\%$}} & \pbox{20cm}{3.43 \\ \small{$p=6\%$}} & \pbox{20cm}{4.26 \\ \small{$p=4\%$}} \\[0.35cm]
\pbox{20cm}{\texttt{Plik} (reference):  Self-interactions} & \pbox{20cm}{4.87 \\ \small{$p=3\%$}} & \pbox{20cm}{0.76 \\ \small{$p=38\%$}} & \pbox{20cm}{2.71 \\ \small{$p=10\%$}} \\%[0.25cm]
\midrule &&&\\[-0.3cm]
\pbox{20cm}{\texttt{CamSpec} (alternative):  $\Lambda$CDM} & \pbox{20cm}{4.82 \\ \small{$p=3\%$}} & \pbox{20cm}{2.01 \\ \small{$p=16\%$}} & \pbox{20cm}{1.96 \\ \small{$p=16\%$}} \\[0.35cm]
\pbox{20cm}{\texttt{CamSpec} (alternative):  Self-interactions} & \pbox{20cm}{2.06 \\ \small{$p=15\%$}} & \pbox{20cm}{1.39 \\ \small{$p=24\%$}} & \pbox{20cm}{1.79 \\ \small{$p=18\%$}}    \\
\bottomrule
\end{tabular}
\caption{Significance of the lensing anomaly without and with neutrino long-range self-interactions, for the two official Planck likelihoods (see text for details).
\emph{Introducing long-range interactions removes the significant preference for extra lensing, rendering the CMB data self-consistent with the assumed cosmology.}
}

\label{tab:chisq}
\end{center}
\end{table*}

\section{Methodology and results}
\label{sec:method} 

Below, we quantify how neutrino long-range interactions alleviate the $A_\mathrm{lens}$ anomaly. For this, we use: the temperature and polarization CMB anisotropies from Planck 2018~\cite{Aghanim:2019ame} (at both high and low multipoles with the reference \texttt{Plik} and alternative \texttt{CamSpec} likelihoods, which among other differences use different polarization masks, dust subtraction and calibration) that we refer to as \textbf{TTTEEE$\boldsymbol{+}$lowl}; the Planck CMB lensing reconstruction from the four-point correlation function~\cite{Aghanim:2018eyx}, that we refer to as \textbf{Lensing}; and the BAO observations from 6dFGS~\cite{Beutler:2011hx}, SDSS-DR7 MGS~\cite{Ross:2014qpa}, and BOSS DR12~\cite{Alam:2016hwk}, that we refer to as \textbf{BAO}. 

We compute cosmological observables by varying all parameters (that is, the standard $\Lambda$CDM parameters, the lensing parameter $A_{\rm lens}$, the total neutrino mass $\sum m_\nu$, and the interaction strength $g m_\nu/M_\phi$) using the extension of the code  \texttt{CLASS}~\cite{Blas:2011rf,Lesgourgues:2011re} from Ref.~\cite{Esteban:2021ozz} (publicly available at \url{https://github.com/jsalvado/class_public_lrs}). We compute $\Delta \chi^2_\mathrm{eff}$ by modifying the \texttt{MontePython}~\cite{Brinckmann:2018cvx} code, and we carry out all our minimizations with the \texttt{BOBYQA} minimizer~\cite{bobyqa} and its Python implementation~\cite{10.1145/3338517}. This minimizer is robust against non-smooth likelihoods, making it ideal for the problem studied here~\cite{Planck:2013pxb, Prasad:2014wva}.

Figure~\ref{fig:triangle} shows that neutrino long-range interactions dilute the preference towards extra lensing. We show the Bayesian $90\%$ credible regions in two-parameter spaces of $\sum m_\nu$, $H_0$, $\sigma_8$, $A_{\rm lens}$ and $g m_\nu/M_\phi$; for the $\Lambda$CDM plus $A_{\rm lens}$ model and for the interacting long-range scenario after marginalizing over all other parameters. We combine Planck temperature, polarization and lensing data with BAO measurements; and we perform a Bayesian MCMC sampling that we have checked to converge as all Gelman-Rubin parameters satisfy $R-1 < 0.02$. In the hatched region, cosmological neutrinos would still be relativistic today, with potential impact on other observables depending on the flavor structure of the interaction~\cite{Esteban:2021ozz}. This figure illustrates the shift induced in the different parameters by the presence of a long range force: the value of $A_{\rm lens}$ is lowered, as CMB lensing is reduced in the presence of these long range interactions. As the interaction strength increases, the parameter $A_{\rm lens}$ becomes more consistent with the canonical $A_{\rm lens}=1$. In this model the cosmological neutrino mass bound is largely relaxed. The other cosmological parameters suffer a mild shift in their best-fit values. In addition, the $H_0$ and $\sigma_8$ tensions are not worsened. 

Precisely quantifying the significance of the lensing anomaly through \cref{fig:triangle}  is subject to prior dependence and Bayesian volume effects. There is infinite parameter space where long-range interactions are degenerate with $\Lambda$CDM: for very large $g m_\nu/M_\phi$, where neutrinos behave as a massless ideal gas (see \cref{fig:w}); for small $g m_\nu/M_\phi$, where the interaction is negligible; and for small $\sum m_\nu$, where interaction effects turn on very late (see \cref{fig:w}). Thus, Bayesian inferences that integrate over $\sum m_\nu$ and/or $g/M_\phi$ will receive large contributions from parameter regions with no new effects and large $A_\mathrm{lens}$. Even if some parameter region with beyond-$\Lambda$CDM effects and $A_\mathrm{lens} \sim 1$ provided an equally good fit  ---hence alleviating the anomaly---, its small relative volume would bias conclusions towards no new effects and a significant lensing anomaly. Moreover, the relative volume of these regions depends on the chosen priors (e.g., linear vs logarithmic) and parameter limits.

Hence, to draw our quantitative conclusions we follow the Planck collaboration~\cite{Planck:2018vyg} and we quantify the significance of the lensing anomaly by computing
\begin{equation}
    \Delta \chi^2_\mathrm{eff} = \chi^2_\mathrm{eff}(A_\mathrm{lens}=1) - \chi^2_\mathrm{eff}(A_\mathrm{lens}\neq 1)  \, ,
    \label{eq:delta_chisq}
\end{equation}
where $\chi^2_\mathrm{eff}$ is the value of $\chi^2$ after minimizing over all cosmological and nuisance parameters. A larger $\Delta \chi^2_\mathrm{eff}$ corresponds to a more significant anomaly. As no integral is performed, this figure of merit avoids any Bayesian volume effect.

\Cref{tab:chisq} shows our main results. As described above, we compute $\Delta \chi^2_\mathrm{eff}$ for different datasets and for the two Planck likelihoods, the reference \texttt{Plik} and the alternative \texttt{CamSpec}; for $\Lambda$CDM and with long-range neutrino interactions. In the former case, we follow the Planck analysis and assume $\sum m_\nu = 0.06$ eV~\cite{Planck:2018vyg}. In the latter, both the total neutrino mass $\sum m_\nu$ and the interaction strength $g m_\nu/M_\phi$ are free parameters to be determined by data. We also show the frequentist $p-$values for excluding $A_\mathrm{lens} = 1$. (Below, we further quantify the tension with the Akaike Information Criterion.)

We first note that when no long-range forces are present in the neutrino sector, the fit to CMB temperature and polarization anisotropy data greatly improves if we introduce the extra parameter $A_{\rm{lens}}$. From a frequentist point of view, there is a 3$\sigma$ preference for $A_{\rm{lens}}\neq 1$ if we use the \texttt{Plik} likelihood, and somewhat smaller for \texttt{Camspec}. Interestingly, this preference drops drastically when considering additional data sets, indicating some internal inconsistency.

The preference for $A_\mathrm{lens} \neq 1$ dilutes when we include neutrino long-range interactions. Its statistical significance merely reaches the $\sim 1.5\sigma$--$2\sigma$ level (or $< 1 \sigma$ if we consider Planck temperature, polarization and lensing data), and therefore it can be interpreted as a simple statistical fluctuation.  Furthermore, the values of $\Delta \chi^2_\mathrm{eff}$ become more uniform and consistent among the different datasets. This is unlike other proposed solutions to the lensing anomaly such as curvature, where BAO data fully restores the preference towards ${A_\mathrm{lens} \neq 1}$.

In order to further assess the impact of our results, we also use one of the information criteria widely exploited
in astrophysical and cosmological studies to discriminate between competing models (see e.g., Refs.~\cite{Liddle:2007fy,Trotta:2008qt}): the frequentist Akaike Information Criterion (AIC)
\begin{equation}
\textrm{AIC}\equiv \chi^2_\mathrm{eff} +2k~,
\end{equation}
where $k$ is the number of free parameters in the model; the second term penalizes models with more free parameters. The best model is the one with the smallest AIC. Following Jeffreys' scale, we rate a difference ${\Delta \text{AIC} > 5}$ or $>10$ as strong or decisive evidence, respectively, against the disfavored model~\cite{Liddle:2007fy}.

This statistical criterion also reflects that neutrino long-range interactions dilute the lensing anomaly. 
Focusing on the TTTEEE+lowl dataset and the \texttt{Plik} likelihood, $\Delta$AIC $=7.66$, and therefore $A_{\rm{lens}}=1$ is penalized in a strong-to-decisive manner with respect to the case in which the lensing amplitude is a free parameter. For the very same data set and likelihood, ${\Delta \text{AIC} < 5}$ when long-range interactions are switched on, and therefore no significant claim can be stated concerning a preference for $A_{\rm{lens}}\neq 1$. Similar results hold for other datasets and the \texttt{CamSpec} likelihood. A non-minimal neutrino sector thus dilutes the preference for extra artificial lensing, making CMB data consistent with the assumed cosmological history of the Universe. 

\begin{table}[hbtp]
\begin{center}
\begin{tabular}{ccc@{\hskip 0.35cm}c@{\hskip 0.35cm}c}
\toprule
& & \pbox{20cm}{TTTEEE\\+lowl} & \pbox{20cm}{TTTEEE\\+lowl \\+lensing} & \pbox{20cm}{TTTEEE\\+lowl \\+ lensing\\+BAO}\\
\midrule
\multirow{2}{*}{\pbox{20cm}{\texttt{Plik} \\(reference)} } & $\sum m_\nu$ [eV] & 0.8 & 2.9 & 0.08 \\
 & $g \, m_\nu/ M_\phi$ & 13 & 317 & $2 \cdot 10^3$ \\
\midrule
\multirow{2}{*}{\pbox{20cm}{\texttt{CamSpec} \\ (alternative)} } & $\sum m_\nu$ [eV] & 0.7 & 0.07 & 0.06 \\
 & $g \, m_\nu/ M_\phi$ & 16 & 7 & 15 \\
\bottomrule
\end{tabular}
\caption{Neutrino mass and long-range interaction strength preferred by different datasets. These values alleviate the lensing anomaly.}
\label{tab:table_BFP}
\end{center}
\end{table}

\Cref{tab:table_BFP} shows the neutrino parameters that  alleviate the lensing anomaly. We show the best-fit values for the total neutrino mass $\sum m_\nu$ and the long-range interaction strength $g m_\nu /M_\phi$ assuming a canonical lensing amplitude $A_\mathrm{lens}=1$,  for both likelihoods and for the three possible data sets here exploited. We note that, in general, CMB data prefers moderate interaction rates $g m_\nu / M_\phi \sim 10$--$10^3$. The preferred neutrino masses, still lying in the sub-eV region compatible with laboratory data~\cite{KATRIN:2019yun}, typically exceed the current cosmological limits for non-interacting neutrinos. More precise measurements of the neutrino mass in the laboratory, which are more model-independent~\cite{Esteban:2021ozz}, could thus shed complementary light onto these models and their early Universe phenomenology. Although the precise best fit changes from one dataset to another, we notice from the contours in Fig.~\ref{fig:triangle} that a wide parameter range points towards $A_\mathrm{lens}=1$. No fine-tuning of the best-fit values is needed to alleviate the $A_{\rm lens}$ tension.

%%%%%%%%%%%%%%%%%%%%%%%%%%%%%%% 
\section{Conclusions and Future Prospects} 
\label{sec:concl} 
The  $\Lambda$CDM picture has provided an extremely good explanation of present cosmological observations. Nevertheless, it relies on very simplistic assumptions on the different ingredients, reflecting our ignorance on the microphysics of the different components of the Universe. Namely, dark energy is simply described by a cosmological constant representing the vacuum energy, dark matter is merely assumed to be a non-interacting dust, and neutrinos are  relativistic species that transition into dust as an ideal gas.

Deviations from this simple and economical scenario are thus natural as data gets more precise. These may show up in self-consistency tests that start out as anomalies and then evolve into incontrovertible evidence pointing to either uncontrolled uncertainties or new physics. The fact that cosmology probes extreme densities also implies that phenomena beyond the Standard Model of Particle Physics could first appear there.

In this work, we have explored if the simplistic and minimal $\Lambda$CDM scenario could alleviate some of its current anomalies if the neutrino sector is richer, leading to unexpected effects. This is somewhat expected, as neutrinos have provided so far the only laboratory evidence for physics beyond the Standard Model~\cite{Pontecorvo:1967fh,Gribov:1968kq,Ahmad:2002jz, Ahmad:2002ka, Ahmed:2003kj,
  Aharmim:2005gt, BeckerSzendy:1992hq, Fukuda:1994mc, Fukuda:1998mi}. 

An especially rich cosmological observable  is the CMB lensing. Neutrinos substantially affect it, to the extent that present and future CMB constraints on neutrino masses strongly rely on measuring it~\cite{TopicalConvenersKNAbazajianJECarlstromATLee:2013bxd}. And, as we have shown, deviations from the minimal ideal gas scheme notably modify the canonical expectations.

A non-ideal neutrino equation of state thus \emph{naturally} provides a physical framework where to solve the so-called lensing ($A_{\rm{lens}}$) anomaly. We have shown that the presence of a long-range interaction in the neutrino sector makes CMB observations consistent with the canonical value $A_{\rm{lens}}=1$. In addition, the interaction strength required to decrease the lensing anomaly down to a statistical fluctuation is consistent with other existing bounds.

Our scenario can be easily explored with other cosmological observations. Future CMB measurements will have superb precision, and they could discriminate between neutrino interactions and uniformly enhanced lensing~\cite{CMB-S4:2016ple,Renzi:2017cbg}. Furthermore, current surveys are rapidly improving our precision of the matter power spectrum~\cite{DESI:2019jxc,DES:2021wwk}, to the extent that they could start being sensitive to effects induced by neutrino self-interactions~\cite{Esteban:2021ozz}. In Ref.~\cite{Esteban:2021ozz} it was explicitly shown that long range neutrino interactions remove the power spectrum enhancement at scales $k \sim 10^{-3}h$/Mpc due to non-relativistic neutrinos falling in the dark matter gravitational wells and contributing to structure growth. As shown there, the future EUCLID survey will significantly improve the current sensitivity to long-range interacting neutrino scenarios. The light new scalar responsible for the long-range interaction could also give rise to observable departures from standard cosmology~\cite{Fardon:2003eh,Kaplan:2004dq,Gu:2003er,Peccei:2004sz,Bean:2000zm,Bean:2001ys}.

More interestingly, due to the direct link to particle physics, neutrino long-range interactions can leave imprints on astrophysical and laboratory probes. Environments with a large neutrino density such as the Sun or core-collapse supernovae might get affected~\cite{Cirelli:2005sg}, and the precise data that future neutrino observatories will gather opens exciting prospects~\cite{Abe:2011ts, Hyper-Kamiokande:2018ofw, SNO:2015wyx, Capozzi:2018dat,Hyper-Kamiokande:2021frf,DUNE:2020ypp,JUNO:2015zny}. Our scenario also predicts that the entire cosmic neutrino background should be condensed in relatively dense lumps that could give additional signals, for instance through their effect on astrophysical neutrinos. Furthermore, in minimal realizations of our model laboratory neutrino mass constraints are not affected~\cite{Esteban:2021ozz}. Thus, the relatively large neutrino masses required to alleviate the lensing anomaly could generate unexpected signals in beta decay~\cite{KATRIN:2001ttj,KATRIN:2019yun} or neutrinoless double-beta decay~\cite{Giuliani:2019uno} searches. Additionally, variations of our model could also induce neutrino decay~\cite{Berezhiani:1991vk,Beacom:2002cb,Barger:1998xk,Fogli:1999qt,Barger:1999bg,Choubey:2006aq,Palomares-Ruiz:2005zbh,Dentler:2019dhz,Lindner:2001th,Kachelriess:2000qc,Ando:2003ie,Fogli:2006xy,Hannestad:2005ex,Maltoni:2008jr,Escudero:2020ped,Escudero:2019gfk}.

Finally, we have only considered the phenomenology induced by the minimal Lagrangian~\eqref{eq:Lagrangian}. Electroweak gauge invariance, or a mechanism that ensures the lightness of the mediator, can only be achieved in a UV-complete model with extra ingredients (see, e.g., Refs.~\cite{Blinov:2019gcj,Blum:2014ewa,Berryman:2018ogk,Kelly:2020pcy}). This may bring up additional phenomenology in other cosmological, astrophysical, or laboratory observations.

We are living exciting times from cosmology. The $\Lambda$CDM paradigm will be profoundly scrutinized, and any inconsistency will shed light onto the most elusive part of our Universe. Neutrinos, ordinarily ghostly particles, may be the path to new cosmological models, with intriguing links to astrophysics and particle physics observations.

%%%%%%%%%%%%%%%%%%%%%% 
\begin{acknowledgments}

We are grateful for helpful discussions with John Beacom, Miguel Escudero, Chris Hirata and Gabriel Vasquez. OM is supported by the Spanish grants FPA2017-85985-P, PROMETEO/2019/083, PID2020-113644GB-I00 and by the European ITN project HIDDeN (H2020-MSCA-ITN-2019//860881-HIDDeN).
JS is supported from the European ITN project H2020-MSCAITN-2019/860881-HIDDeN, the Spanish grants PID2019-108122GBC32, PID2019-105614GB-C21, and to the Unit of Excellence María de Maeztu 2020-2023 (CEX2019-000918-M).
\end{acknowledgments}

\bibliography{biblio}
\end{document}